\documentclass[twocolumn,preprintnumbers,floatfix,prb,superscriptaddress]{revtex4}
\usepackage{graphicx}
\usepackage{graphicx}
\usepackage{dcolumn} 
\usepackage{bm}
\usepackage{epsfig}
\begin{document} 

\title{Strain effects to optimize the thermoelectric properties of hole-doped La$_2$NiO$_{4+\delta}$ via ab initio calculations}

\author{Victor Pardo}
 \email{victor.pardo@usc.es}
\affiliation{
Departamento de F\'{\i}sica Aplicada, Universidade
de Santiago de Compostela, E-15782 Santiago de Compostela,
Spain
}
\affiliation{
Instituto de Investigaci\'{o}ns Tecnol\'{o}xicas, Universidade
de Santiago de Compostela, E-15782 Santiago de Compostela,
Spain
}

\author{Antia S. Botana}
\affiliation{
Departamento de F\'{\i}sica Aplicada, Universidade
de Santiago de Compostela, E-15782 Santiago de Compostela,
Spain
}
\affiliation{
Instituto de Investigaci\'{o}ns Tecnol\'{o}xicas, Universidade
de Santiago de Compostela, E-15782 Santiago de Compostela,
Spain
}

\author{Daniel Baldomir}
\affiliation{
Departamento de F\'{\i}sica Aplicada, Universidade
de Santiago de Compostela, E-15782 Santiago de Compostela,
Spain
}
\affiliation{
Instituto de Investigaci\'{o}ns Tecnol\'{o}xicas, Universidade
de Santiago de Compostela, E-15782 Santiago de Compostela,
Spain
}

\date{\today}

\begin{abstract}

Thermoelectric properties of the system La$_2$NiO$_{4+\delta}$ have been recently discussed [Phys. Rev. B 86, 165114 (2012)] via ab initio calculations. An optimum hole-doping value was obtained with reasonable thermopower and thermoelectric figure of merit being calculated. Here, a large increase in the thermoelectric performance through lattice strain and the corresponding atomic relaxations is predicted. This increase would be experimentally attainable via growth in thin films of the material on top of different substrates. A small tensile strain would produce large thermoelectric figures of merit at high temperatures, $zT$ $\sim$ 1 in the range of oxygen excess $\delta$ $\sim$ 0.05 - 0.10 and in-plane lattice parameter in the range 3.95 - 4.05 \AA. In that relatively wide range of parameters, thermopower values close to 200 $\mu$V/K are obtained. The best performance of this compound is expected to occur in the high temperature limit.

\end{abstract}

\maketitle


\section{Introduction}

The thermoelectric (TE) performance of a material is usually quantified by means of the so-called dimensionless figure of merit $zT$= $\sigma$TS$^2$/$\kappa$, where $S$ is the thermopower, $\sigma$ the electrical conductivity and $\kappa$ the thermal conductivity that can be expressed as the sum of electronic and lattice contributions, $\kappa= \kappa_e+ \kappa_l$. $zT$ $>$ 1 is required for applications,\cite{review_nat_therm,zt_minimum} and in order to maximize $zT$, a high $S$, high $\sigma$ and low $\kappa$ are required. All the electronic magnitudes (electrical conductivity, thermopower and the electronic component of the thermal conductivity) are inter-related and their simultaneous optimization is somehow conflicting. Low-gap semiconductors are the best meeting the compromises considering only the electronic properties. If one reduces (increases) the band gap of a semiconductor, the conductivity increases (decreases) and the Seebeck coefficient goes down (up). Thus, there is a compromise situation at intermediate doping levels where TE efficiency is enhanced. To reduce the lattice thermal conductivity, nanostructuring is a common strategy.\cite{nanostructuring} The situation gets more complicated when instead of standard parabolic-band semiconductors, strongly correlated electron systems with a low band gap are considered.\cite{maekawa} These can in principle be equally effective from an electronic structure point of view, but they provide additional ingredients that allow to tune the band structure in order to optimize separately the different magnitudes involved in the TE response: having localized electrons allows for the tuning of band splittings, band widths, correlation and ordering effects, etc. Moreover, band engineering is possible in strongly correlated electron systems in order to tune the electronic structure for an enhanced TE response, i.e. they are more flexible than standard semiconductors for a tunability of their band structure through  magnitudes other than doping or nanostructuring (commonly used for sp semiconductors). Parameters such as strain, ordering phenomena, electron localization mechanisms, pressure effects, etc. often play a big role in their electronic structure.\cite{review_khomskii} In particular, strain engineering has recently been shown as a direct method to control the TE properties in topological insulator materials Bi$_2$Se$_3$ and Bi$_2$Te$_3$.\cite{bite_se}  

Among oxides and other 3d electron systems, misfit layered cobaltates have drawn significant attention as TE materials\cite{review_nat_therm, naxcoo_singh, prb_ca3co4o9, prb_misfit, jssc} because of the large thermopower observed together with metallic conductivity. Their interesting TE properties stem from the existence of two electronic systems within those cobaltates.\cite{prb_eg_a1g} Due to the trigonally distorted octahedral environment, the Co t$_{2g}$ bands are split between a highly mobile wide band that drives metallic conduction (of e$_g^*$ character) and a narrower band (with a$_{1g}$ parentage) that provides the large Seebeck coefficient found in that family of oxides.\cite{naxcoo2_terasaki} Similar electronic structure and promising TE properties via hole-doping have been predicted in La$_2$NiO$_{4+\delta}$ by means of ab initio calculations.\cite{la2nio4_therm_prb} The existence of localized and itinerant electrons in that system is also at the heart of a reasonable TE response. In that paper, the electronic structure under doping was analyzed in order to understand which bands are influenced by doping and what amount of oxygen excess is needed to enlarge the TE efficiency. Experimentally, the relevant TE parameters observed typically at room temperature are $S$ $\sim $100 $\mu$V/K,\cite{la2nio4_Seebeck_delta} $\rho$ $\sim$ 5 m$\Omega$.cm \cite{la2nio4_films_sigma,la2nio4_sigma_singlecrystals} and  $\kappa$ $\sim$ 7.5 W/m.K,\cite{la2nio4_kappa} which add up to a $zT$ at room temperature of about 0.01. This is comparable to other promising compounds such as CrN.\cite{camilo_apl,camilo_prb,crn_antia}  

Ref. \onlinecite{la2nio4_therm_prb} suggested that the appropriate oxygen content $\delta$ together with a thin film growth could significantly enhance $zT$. Growing thin films introduces the additional ingredient of strain due to lattice size mismatch when the oxide is grown on different substrates. In this paper we will analyze how one can refine even further the TE properties of La$_2$NiO$_{4+\delta}$ when both oxygen-excess and strain are introduced.  We will perform a band engineering study of how these strain effects can enhance the TE performance in a 3d electron system that has been predicted in the past to yield an interesting TE response. The goal of this work would be designing the proper growth conditions for an improved TE efficiency. 

The paper is organized as follows: Section \ref{compdet} will present the calculation details, Section \ref{results} will describe the electronic structure and its strain dependence, analyze the TE properties calculated for various oxygen contents, and finally we will summarize the main conclusions of the work in Section \ref{summary}.

\section{Computational Procedures}\label{compdet}

Our electronic structure calculations were  performed within density functional
theory\cite{dft,dft_2} using the all-electron, full potential code {\sc wien2k}\cite{wien}
based on the augmented plane wave plus local orbitals (APW+lo) basis set.\cite{sjo}
For the exchange-correlation functional in the structural relaxations we have used the Perdew-Burke-Ernzerhof version of the generalized gradient approximation\cite{gga} (GGA). To study the effects of strain, we have analyzed different in-plane a,b lattice parameters, and for each of them the c parameter was calculated. Consequences will be drawn for strains in thin films but the actual calculations are done in bulk-like unit cells with tetragonal symmetry for various c/a ratios. 


For the calculations of the transport properties, we have used the recently developed Tran-Blaha modified Becke-Johnson (TB-mBJGGA) potential.\cite{TBmBJLDA} This has been shown to provide an accurate account of the electronic structure of correlated compounds\cite{crn_antia,sto_arxiv} using a parameter-free description,  without invoking for a particular value of U selected by hand as typically done in the LDA+U method, and also yielding accurate band gaps for most semiconductors.

The transport properties were calculated using a semiclassical solution based on Boltzmann's transport theory within the constant scattering time approximation by means of the {\sc BoltzTrap} code, that uses the energy eigenvalues calculated by the {\sc wien2k} code.\cite{boltztrap} We refer the reader to Ref. \onlinecite{boltztrap} for details on how the different transport coefficients are obtained. In this case, denser k-meshes are required, in our case up to 40 $\times$ 40 $\times$ 15 to reach convergence. The constant scattering time approximation assumes the relaxation time $\tau(\epsilon)$ as energy-independent. This results in expressions of both the thermopower and the TE figure of merit with no dependence on $\tau$ (they can be directly obtained from the band structure without any assumed parameters). This approximation has been used succesfully to describe several potentially useful TE materials.\cite{singh_1, singh_2, singh_3,singh_4,singh_5}

The simulation of different doping levels was performed in two different ways: i) by a shift in the chemical potential for the undoped compound, calculating the hole concentration at each temperature. We have seen in previous works\cite{la2nio4_therm_prb} that some of the essential features and trends can be obtained by this method, specially at extremely low doping,\cite{sto_arxiv} ii) using the virtual crystal approximation (VCA), that provides a more accurate description of the doped material and yields substantial differences in quantitative values of the TE properties, as we will discuss in detail below. Both calculations treat the dopants in an average way, whereas the actual oxygen-excess atoms will reside in some particular positions in the lattice and can lead to polaronic effects, dopant orderings, etc. that will be partly missed by our calculations. In the low-doping regime our calculations will be more reliable.

All calculations were fully converged with respect to all the parameters used. In particular, we used R$_{mt}$K$_{max}$= 7.0, a k-mesh of 10 $\times$ 10 $\times$ 4, and muffin-tin radii of 2.35 a.u. for La, 1.97 a.u. for Ni and 1.75 a.u. for O.

\section{Results}\label{results}

\subsection{Revisiting the electronic structure and magnetic properties of unstrained La$_2$NiO$_4$}

\begin{figure*}[ht]
\begin{center}
\includegraphics[width=0.32\textwidth,draft=false]{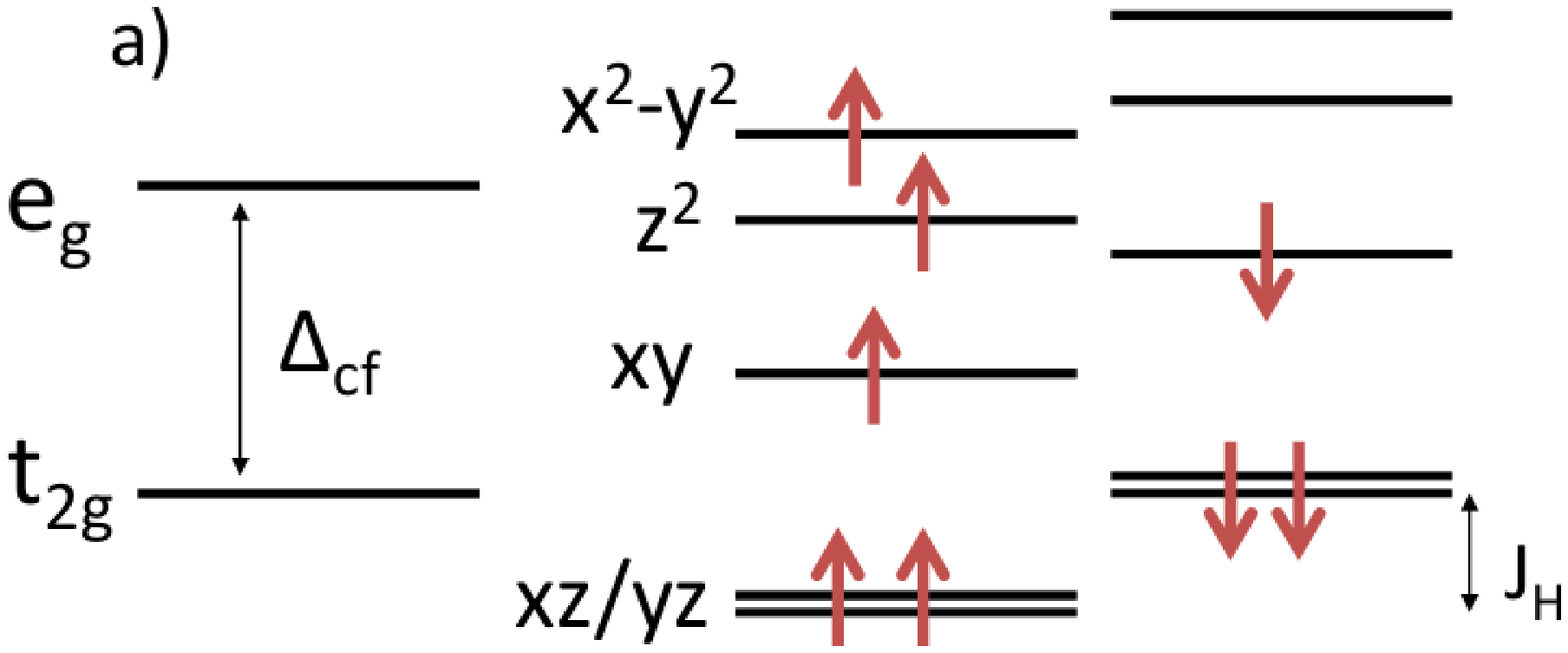}
\includegraphics[width=0.32\textwidth,draft=false]{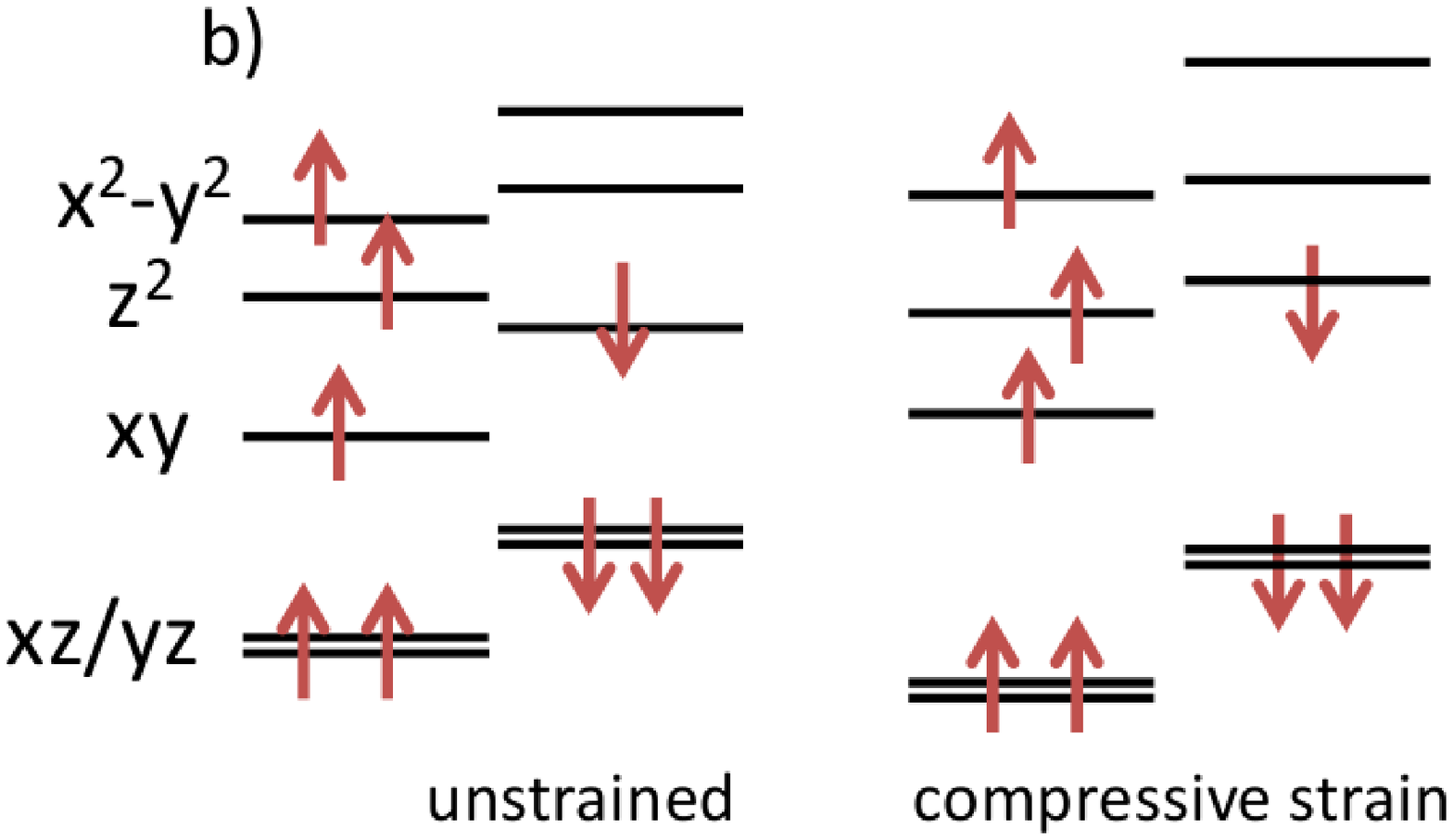}
\includegraphics[width=0.32\textwidth,draft=false]{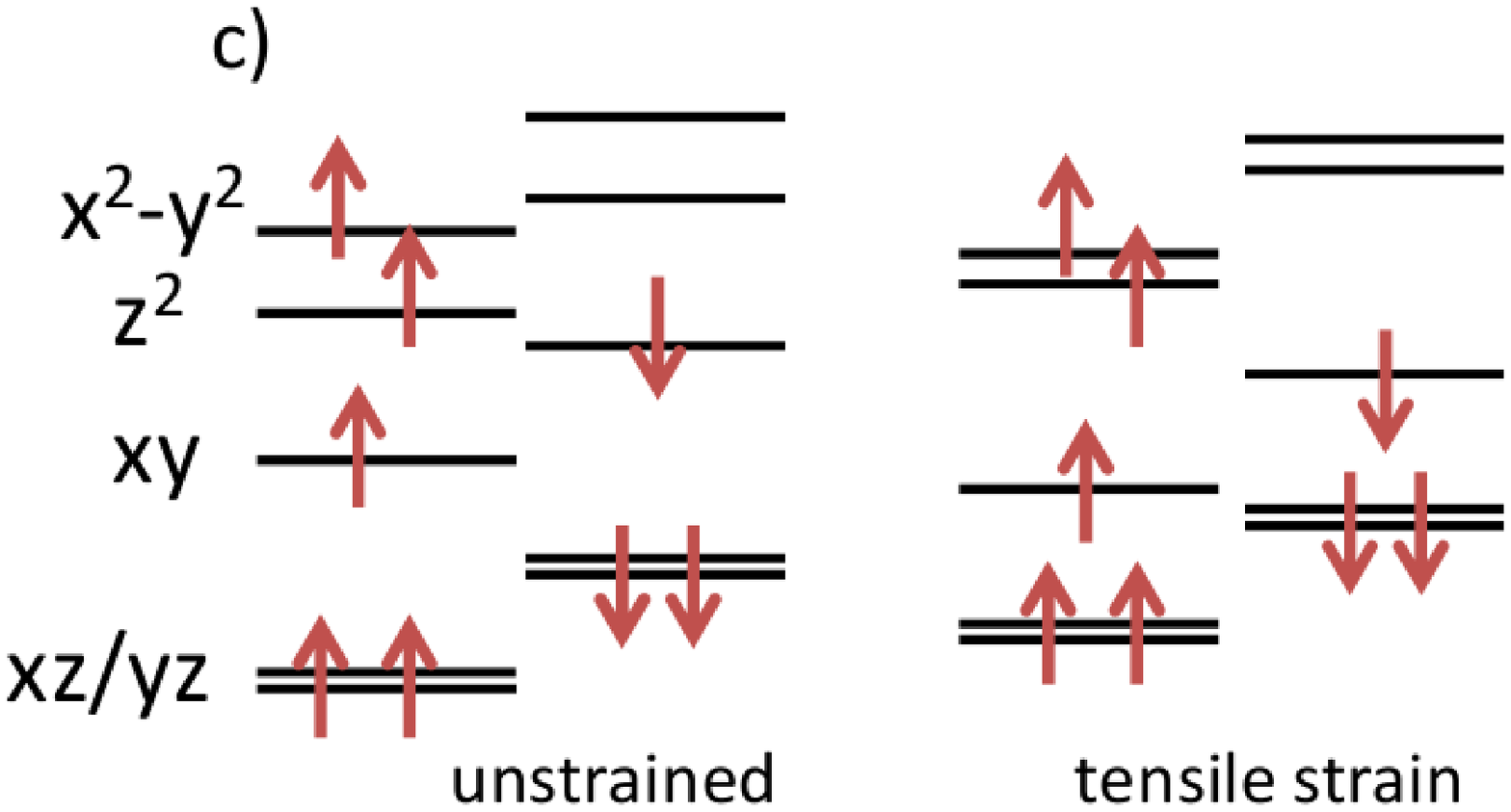}
\caption{(Color online.) Sketch of the crystal field levels of a Ni$^{2+}$:d$^8$ cation in a HS state to observe the effects of: a) a distortion below octahedral symmetry in an elongated octahedron, and the effects of both compressive (b) and tensile (c) strain. The gap is expected to be reduced (increased) for the compressive (tensile) strain case.}\label{cfs}
\end{center}
\end{figure*}

Let us recall some basic ideas about the electronic structure of La$_2$NiO$_{4+\delta}$. This is a compound containing Ni$^{2+}$:d$^8$ cations in a largely elongated octahedral environment. The crystal field splittings in such geometry lead to a breakup of the e$_g$ degeneracy with the d$_{x^2-y^2}$ band higher in energy than the d$_{z^2}$ (as sketched in Fig. \ref{cfs}a). A significant Hund's rule coupling stabilizes the high-spin state (HSS) t$_{2g}$$^6$e$_g$$^0$ (S= 1). In previous works,\cite{la2nio4_therm_prb,la2nio4_goodenough} it was seen that hole-doping via oxygen exccess introduces holes in the Ni d$_{x^2-y^2}$ band, which is somewhat localized. This $d_{x^2-y^2}$ band below the Fermi level can be substantially modified in a process one can call ``band engineering'' in order to improve the TE response of the compound. To that end, one needs e.g. a larger thermopower (proportional to the derivative of the density of states (DOS), and hence increasing with a reduced bandwidth) without compromising the electrical conductivity. One possibility for that is to bring the d$_{z^2}$ band (whose bandwidth is smaller due to the comparably smaller off-plane hopping)  closer to the Fermi level, and make use of its reduced band width to enlarge the thermopower. Of course, this is not exactly so because, as mentioned above, the electronic-only magnitudes in semiconductors are all closely related, and in principle a reduced band width might lead to an increased band gap and a reduction of the conductivity. Yet an optimum performance can be sought, as we will see below. In any case, to bring the $d_{z^2}$ higher in energy (closer to the Fermi level for the majority spin channel), this band needs to be destabilized with respect to the $d_{xy}$ (which itself is split from the lower-lying $d_{xz}$/$d_{yz}$ doublet) and $d_{x^2-y^2}$ bands (see Fig. \ref{cfs}). Thus, enlarging the a,b plane could produce the required effect. This would, however, significantly reduce the in-plane conductivity, which is the largest component in this layered compound, with the corresponding fall-out in figure of merit (let us recall, zT=$\sigma$$S^2$T/$\kappa$). 
If, however, a,b lattice parameters are reduced, $\sigma$ can be increased, but the $d_{x^2-y^2}$ band will be moved up in energy so that it will remain the only accessible band via hole-doping (a sketch of how all these bands shift with strain can be seen in Fig. \ref{cfs}). In that case, the band will become broadened (lower peaks for the DOS, smaller thermopower) by the increased in-plane hopping caused by the smaller lattice parameter. Hence, there will be optimum values (a compromise) for both strain and doping (trying to tune the oxygen excess to reach the chemical potential where thermopower is maximized) that one could use for designing thin films with optimal conditions so that an enhanced TE response can be obtained. Changes in the in-plane lattice parameters are induced by growing epitaxial thin films on different substrates. As explained above, a thin film geometry is beneficial for the overall TE response since $\kappa$ will be drastically reduced and $\sigma$ has been observed to increase due to its large two-dimensional anisotropy.\cite{la2nio4_films_sigma}

Concerning the magnetic order, it has been shown that La$_2$NiO$_4$ (with Ni$^{2+}$:d$^8$ S=1, HSS) is an antiferromagnet with an in-plane ordering such that an antiferromagnetic (AF) interaction between nearest neighbor Ni atoms is stabilized through an e$_g$$^2$-O-e$_g$$^2$ superexchange interaction. The commensurability of the ordering has been, however, put into question.\cite{la2nio4_magn_short_range} Careful studies with respect to the oxygen content show how the N\'eel temperature is significant even at values of $\delta$ $\sim$ 0.1,\cite{la2nio4_TN_delta} and could vanish at $\delta$ $\sim$ 0.14.\cite{la2nio4_Seebeck_delta} We will focus our calculations in that doping level region ($\delta$ $\leq$ 0.15), where magnetic order exists. All our calculations assume an in-plane checkerboard AF ordered phase.

\subsection{Changes in the electronic structure caused by strain}

We have performed calculations at various in-plane lattice parameters, that try to simulate both compressive and tensile strains caused by growth of La$_2$NiO$_{4+\delta}$ thin films on top of different substrates. Calculations are carried out in bulk-like unit cells but help us draw conclusions about the changes in the electronic structure undergone by the compound if grown epitaxially on top of different substrates, with different in-plane strains and c/a ratios. 

We have optimized the out-of-plane c lattice parameter of the structure for each value of the in-plane a, b lattice parameters chosen. We have performed GGA calculations for several values of the in-plane lattice parameters from 3.7 to 4.1 \AA. The bulk value of the $a$ lattice parameter in the tetragonal structure\cite{struct_tetragonal} of La$_2$NiO$_4$ is 3.89 \AA. For each a, c combination, the atomic positions were relaxed and the minimum-energy c-value was obtained for each in-plane area. In Table \ref{latt_param} we can see the results of the lattice parameter optimizations. Seeing the bulk value of the lattice parameter for La$_2$NiO$_4$, it is clear that when grown on a SrTiO$_3$ substrate, it would be roughly unstrained. Films have been grown in the past\cite{la2nio4_films_sigma} on NdGaO$_3$ (with slightly lower lattice parameter, somewhat similar to LaAlO$_3$). Other substrates can be found with different lattice parameters, we have tried to study the evolution of the properties with strain giving an interval of a,b in-plane lattice parameters that covers most of the typically available substrates. Probably the extreme cases we have considered will not be stable in the laboratory, but still help us drawing conclusions about the evolution of the transport properties with strain.

\begin{table}[h!]
\caption{In-plane vs. out-of-plane optimized lattice parameters for various strains. A compression in the xy-plane leads to an elongation in c to try to keep the volume constant.}\label{latt_param}
\begin{center}
\begin{tabular}{|c|c|}
\hline
a (\AA) & c (\AA) \\
\hline
\hline
3.74 & 9.39 \\
3.85 & 8.98 \\
3.93 & 8.85 \\
4.00 & 8.72 \\
4.11 & 8.51 \\
\hline
\end{tabular}
\end{center} 
\end{table}

\begin{figure*}[ht]
\begin{center}
\includegraphics[width=0.49\textwidth,draft=false]{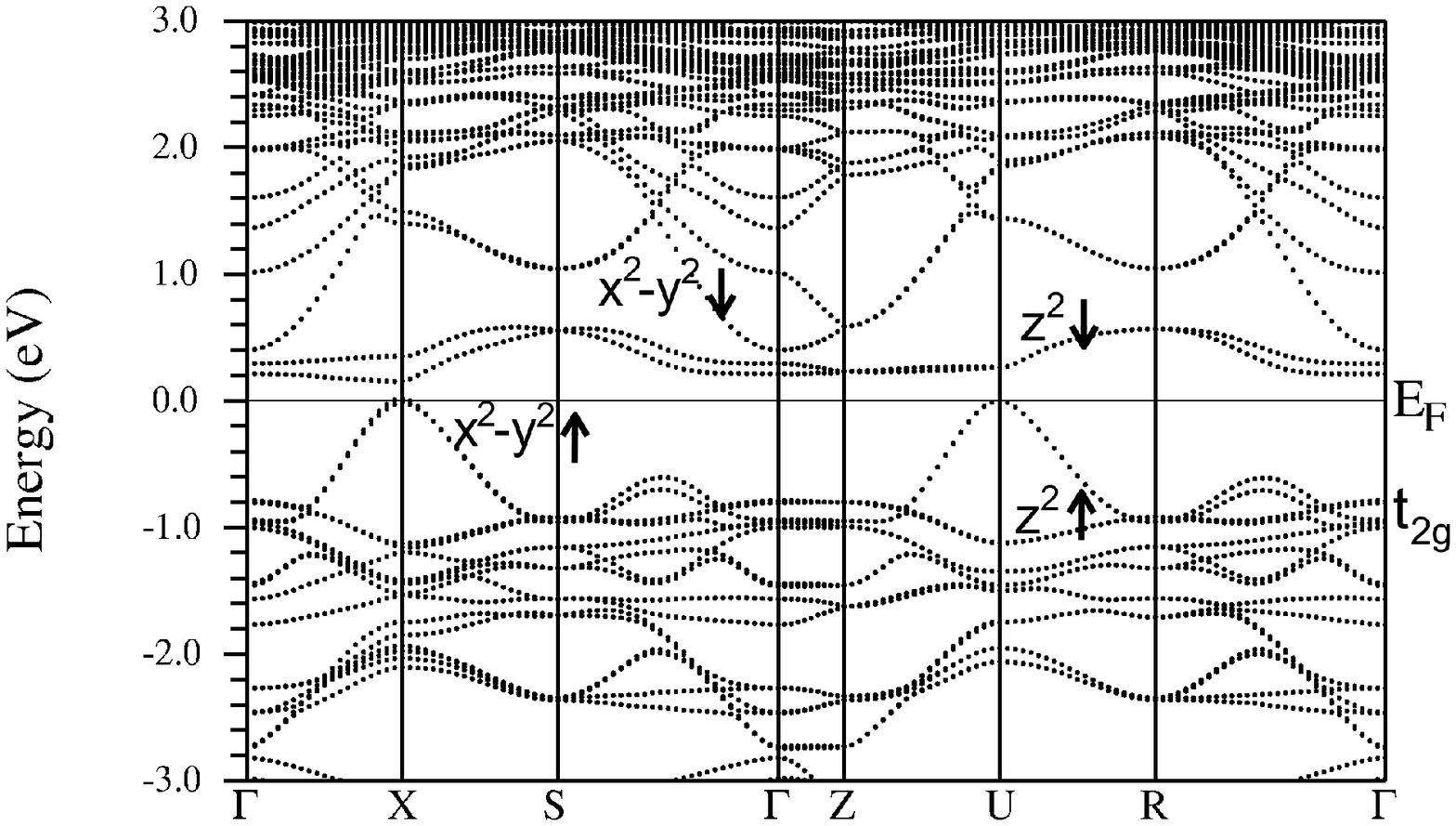}
\includegraphics[width=0.49\textwidth,draft=false]{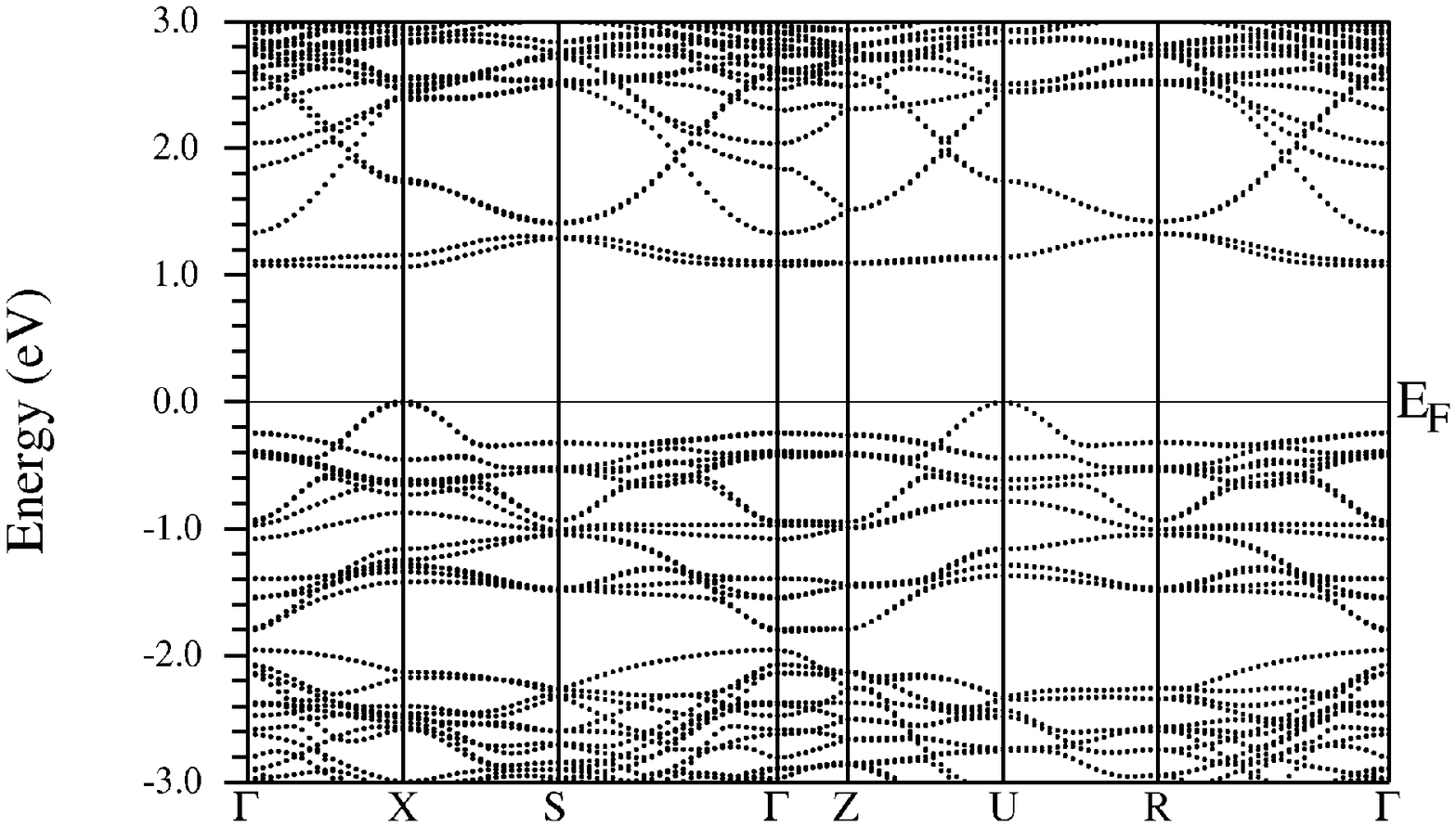}
\caption{(Color online.)  Band structure (calculated within GGA) for two types of strain: compressive on the left (a= 3.74 \AA) and tensile on the right (a= 4.11 \AA). Notice the displacement of the d$_{x^2-y^2}$ band (easy to identify with maxima at X and U, being 1 eV wide, and occuring just below the Fermi level specially in the left panel) to a lower position in the tensile strain case. A significant reduction of the gap occurs for the compressive strain case and a somewhat smaller band width of the d$_{x^2-y^2}$ band is obtained in the tensile strain case.}\label{bs_strain}
\end{center}
\end{figure*}

We know from previous studies and the discussions above that the only band available in unstrained La$_2$NiO$_{4+\delta}$ to be populated with holes via oxygen excess is the majority-spin d$_{x^2-y^2}$ band. Compressive strain (reducing the in-plane lattice parameter) will move this d$_{x^2-y^2}$ band further up in energy but at the same time it will increase its bandwidth (together, an increase in conductivity and decrease in the derivative of the DOS, and hence a reduction of the thermopower, is expected). If, on the other hand tensile strain is applied, this band moves further down in energy (it gets stabilized by the larger in-plane lattice parameter) closer to the energy window where the majority d$_{z^2}$ band and the minority t$_{2g}$ bands reside. Thus, the effects of applying a tensile strain are slightly more complicated because more bands are involved. Moreover, tensile strain would reduce the in-plane conductivity (due to the larger in-plane lattice parameter) but could enlarge the Seebeck coefficient if a large DOS is retained close to the Fermi level (accessible via hole-doping).

Figure \ref{bs_strain} shows the band structure for the two types of strain studied: compressive and tensile. We see that compressive strain (on the left) reduces significantly the band gap, driving the system towards a bigger in-plane conductivity caused by the reduced Ni-Ni in-plane distance and corresponding increase in the Ni d$_{x^2-y^2}$ bandwidth (larger in-plane hopping mediated by those orbitals via a large $\sigma$-bond with oxygens). The d$_{x^2-y^2}$ band is significantly less wide in the right panel (tensile strain) due to the reduced hopping caused by the enlarged in-plane lattice parameter. Moreover, tensile strain leads to a stabilization of the d$_{x^2-y^2}$ band, that places it closer to all the other occupied bands (we see the band very distinctly below the Fermi level in the compressive strain case, where the opposite is true, but somewhat mixed with many other bands for the tensile one). 

There are various factors at play here that can enhance or reduce the different TE properties: conductivity and thermopower. Thus, calculations need to be performed to account for them all (doping, strain, changes in bandwidth, introduction of more bands near the Fermi level, etc.) properly. All in all, we have various conflicting parameters to tune and calculations are required to see the balance between them in order to obtain a compromise that might enhance the TE response.

\subsection{Influence of strain on the thermoelectric properties}

We can quantify all the important magnitudes for the TE response of a material in terms of the dimensionless TE figure of merit $zT$= $\sigma$TS$^2$/$\kappa$. As mentioned above, $zT$ needs to be larger than 1 for applications. It is often quoted that a thermopower larger than 200 $\mu$V/K is needed for a high-performance TE response. We can, from first principles calculations through the electronic structure of the material, give an estimate of the electronic part of the figure of merit. However, for the thermal conductivity, $\kappa$= $\kappa_{e}$ + $\kappa_{l}$, we can only estimate the electronic part, which becomes more important in the itinerant electron limit. In the localized limit, $\kappa_{l}$ (the component due to the lattice) will be the only important contribution. Our results obtained from electronic structure calculations will be an upper limit for the overall $zT$, but still helpful to understand how to optimize and engineer a better response in this and other related systems.

\begin{figure*}[ht]
\begin{center}
\includegraphics[width=0.32\textwidth,draft=false]{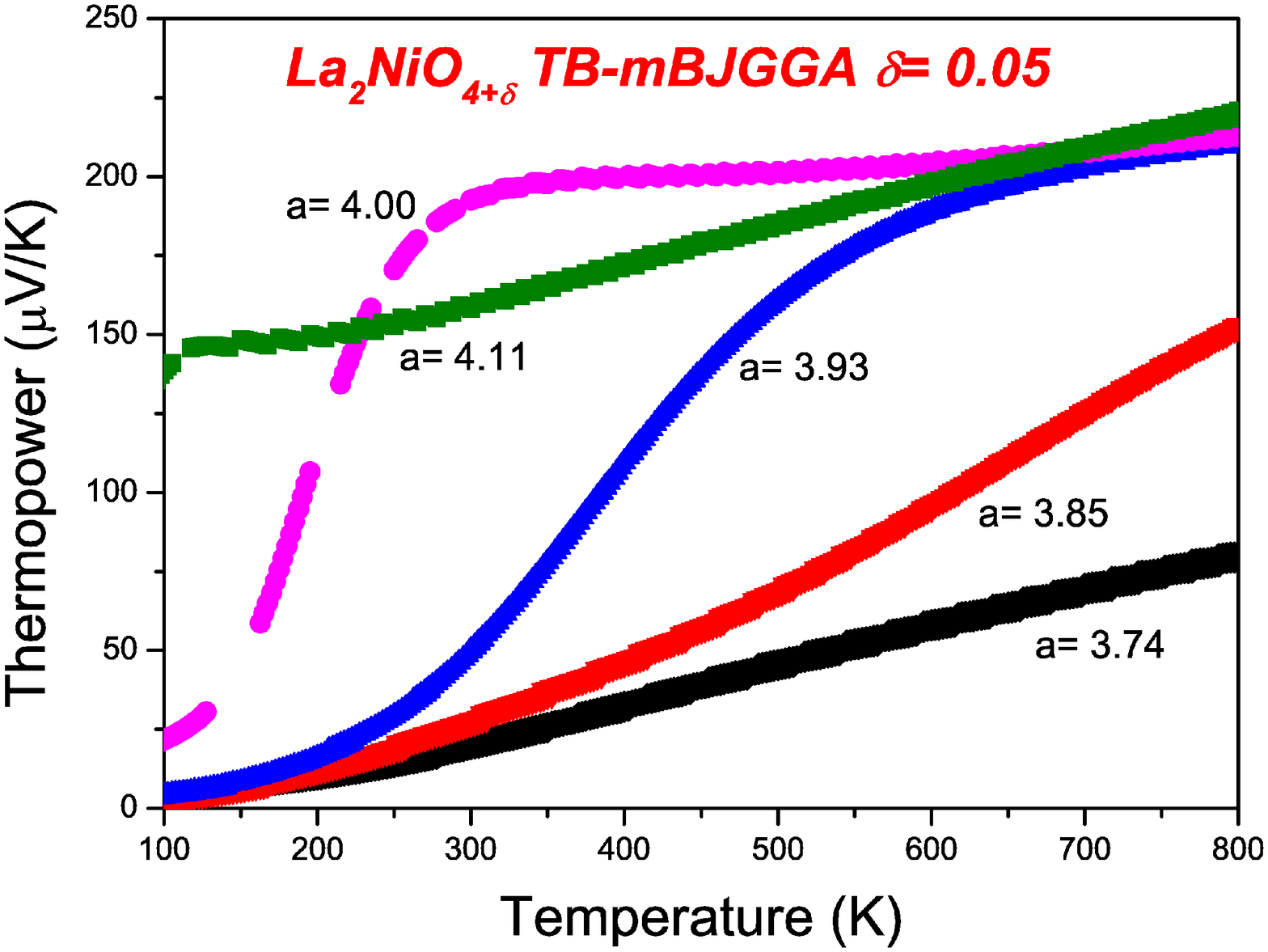}
\includegraphics[width=0.32\textwidth,draft=false]{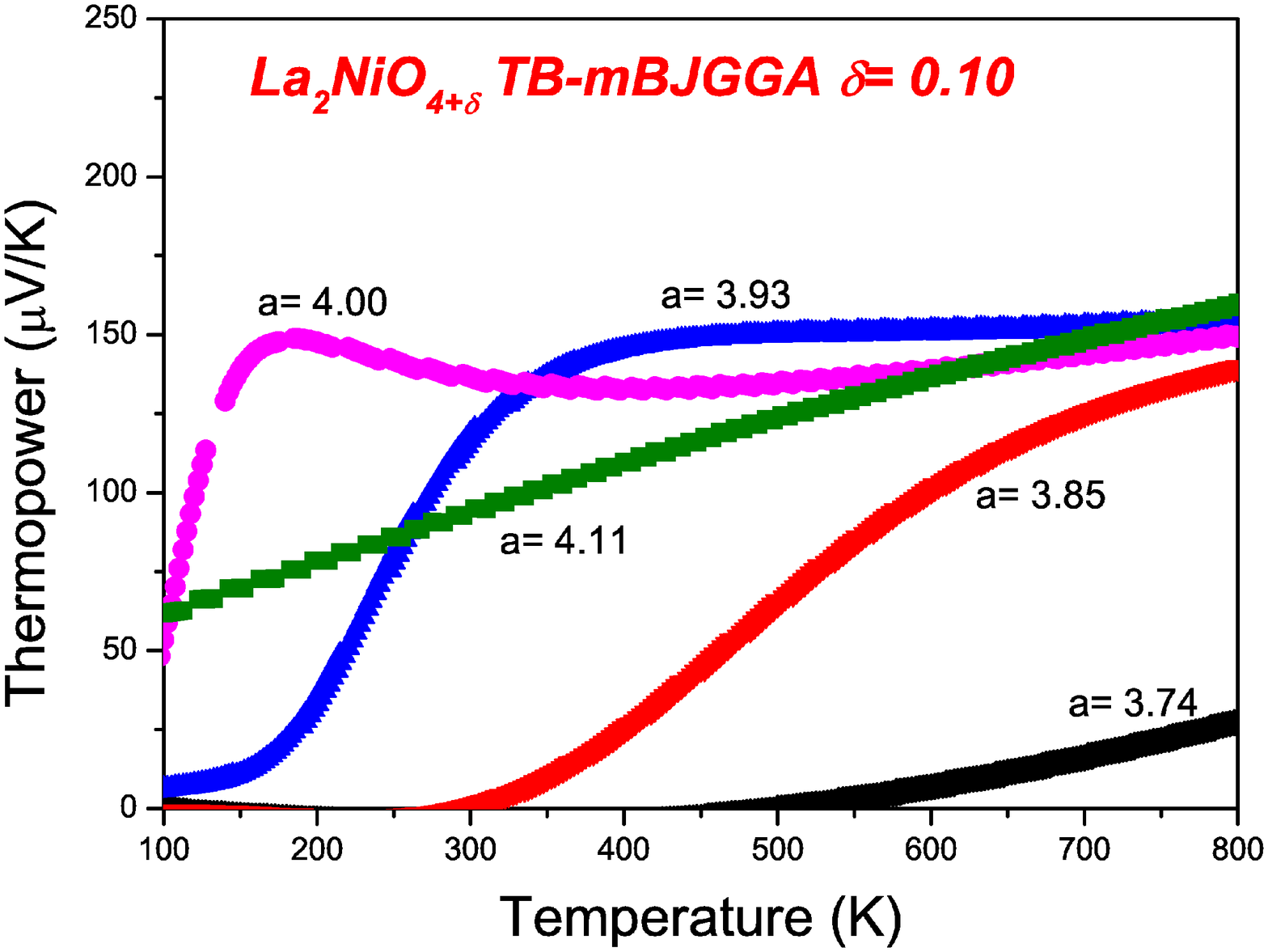}
\includegraphics[width=0.32\textwidth,draft=false]{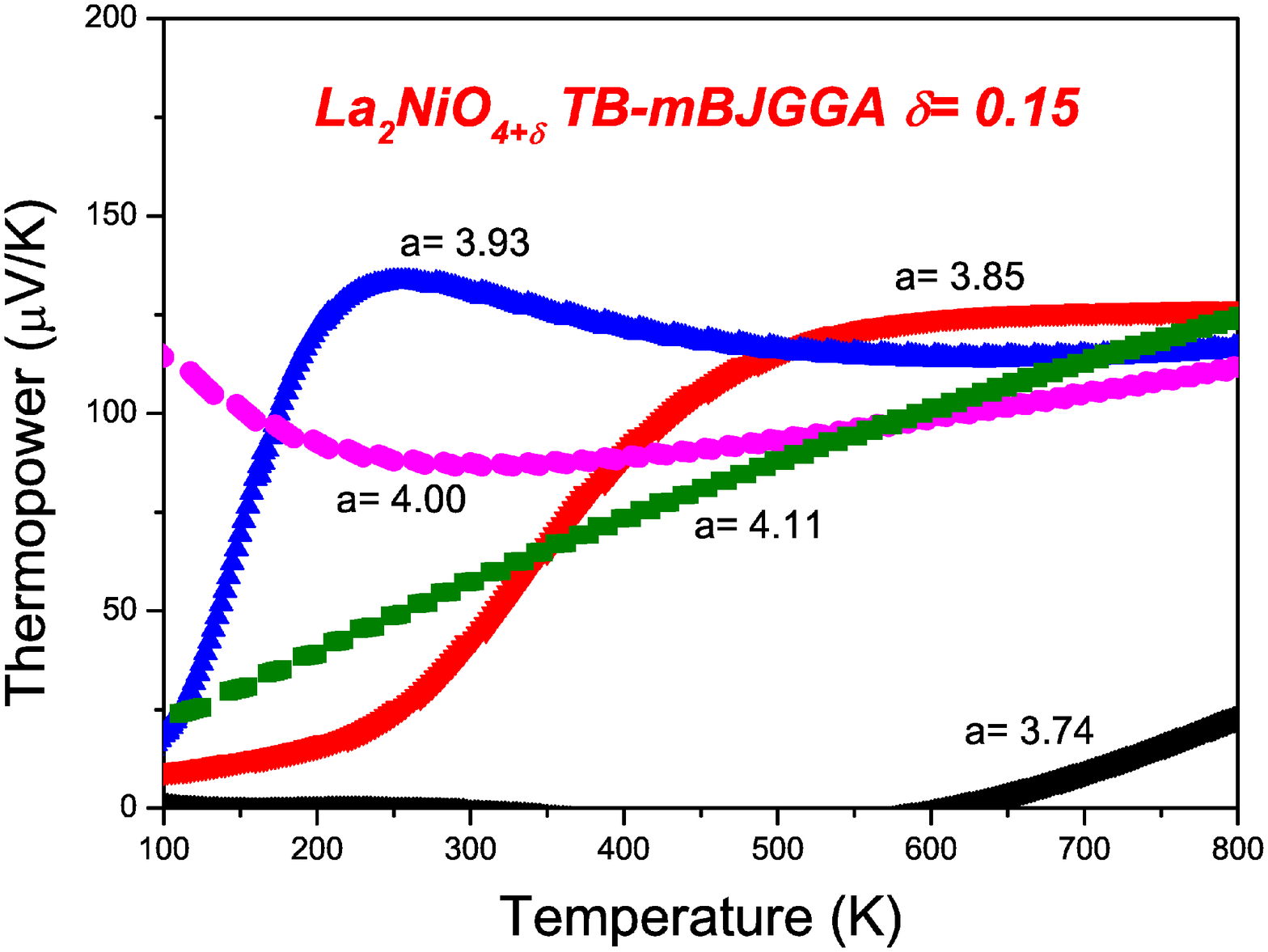}
\includegraphics[width=0.32\textwidth,draft=false]{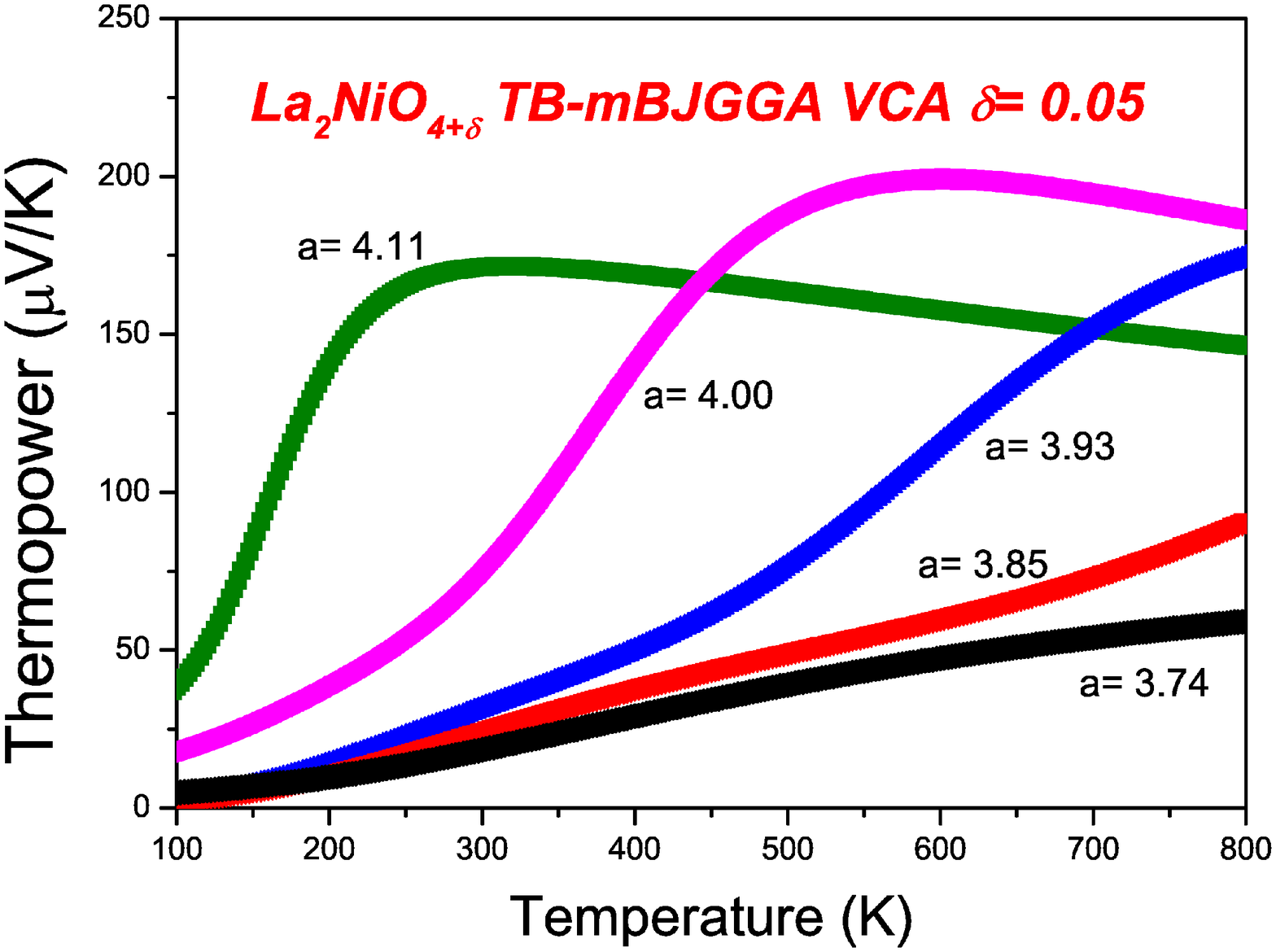}
\includegraphics[width=0.32\textwidth,draft=false]{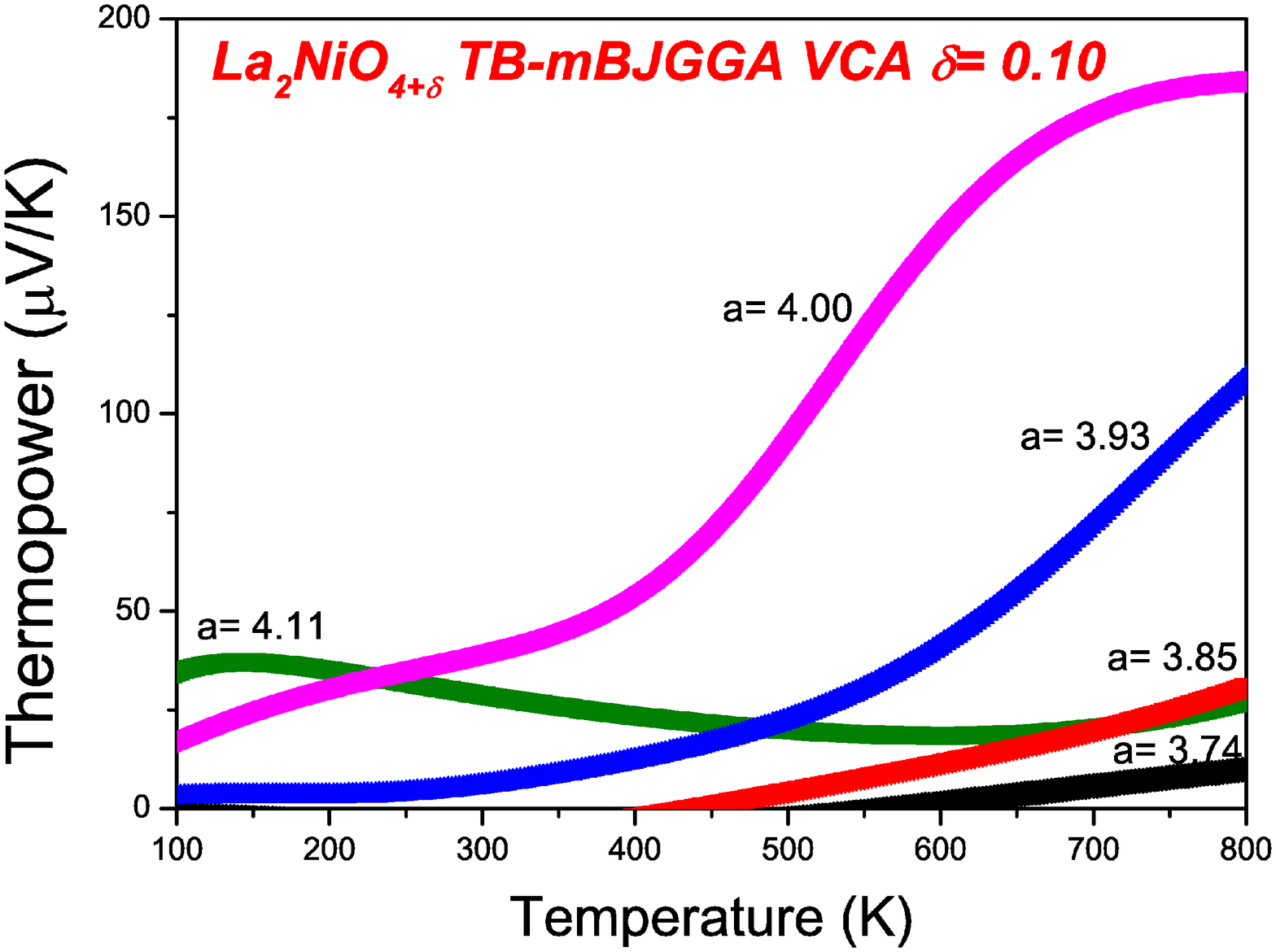}
\includegraphics[width=0.32\textwidth,draft=false]{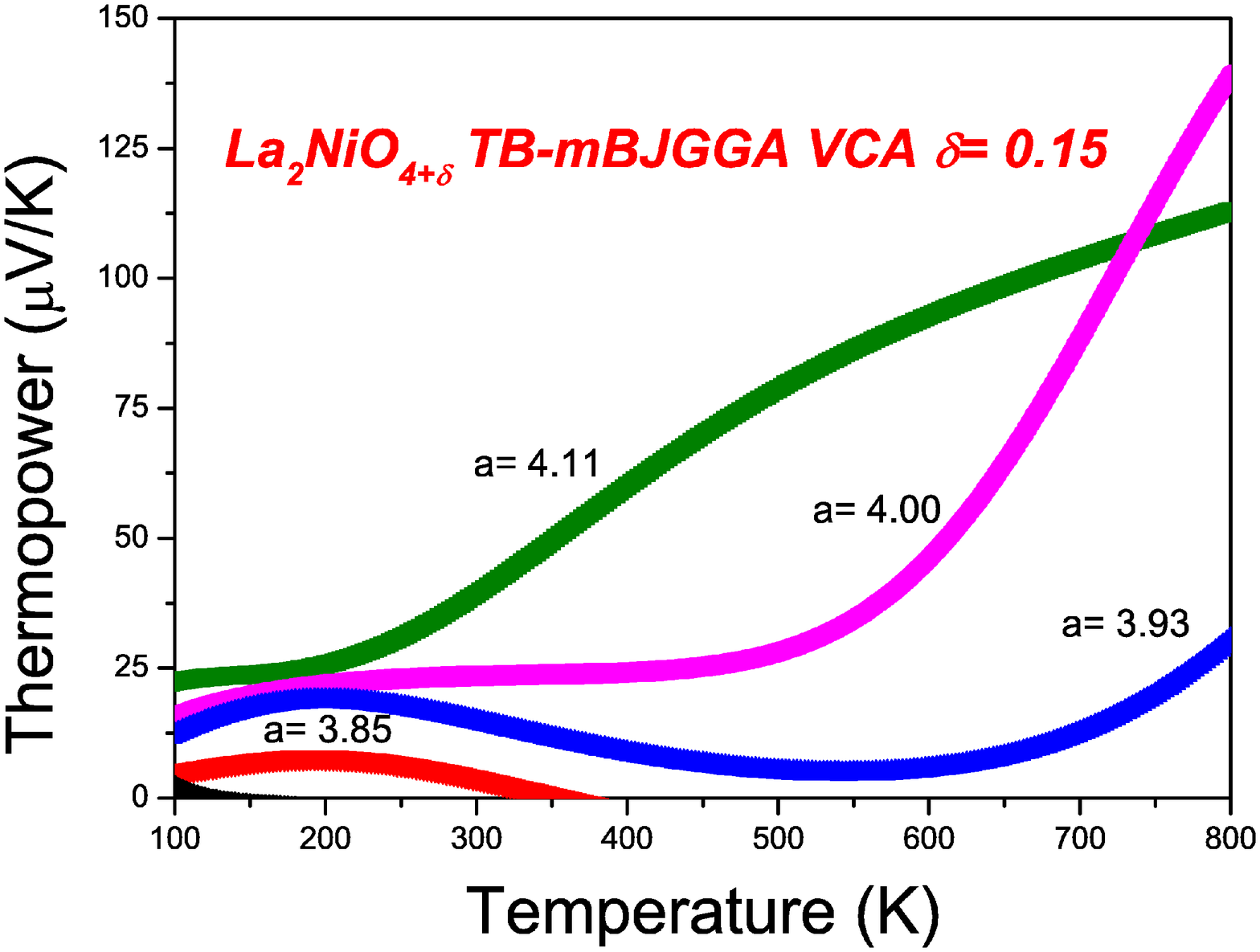}
\caption{(Color online.)  Thermoelectric power of La$_2$NiO$_{4+\delta}$ as a function of temperature for three doping values: $\delta$= 0.05, $\delta$= 0.10 and $\delta$= 0.15 for various values of the in-plane lattice parameter (different strains). The upper panel was calculated for an undoped La$_2$NiO$_4$ shifting the chemical potential to the desired carrier concentration, whereas the lower panel was calculated using the VCA. The latter yields a more reliable and accurate description of the system. The exchange-correlation potential used was TB-mBJGGA. Observe that for various lattice parameters in the doping range $\delta$= 0.05 - 0.10, thermopower exceeding 150 $\mu$V/K are predicted. }\label{S_delta}
\end{center}
\end{figure*}

The first parameter to analyze the TE response of the material is the Seebeck coefficient, that can be calculated independently of the scattering time in the constant scattering time approximation we are using for our calculations. Figure \ref{S_delta} shows the thermopower calculated for three values of hole doping in La$_2$NiO$_{4+\delta}$: $\delta$= 0.05, 0.10 and 0.15, for several values of the lattice parameter from 3.74 to 4.11 \AA, that simulate both the compressive and tensile strain limits (let us recall the unstrained value of the in-plane lattice parameter is 3.89 \AA). Results are presented using VCA (lower panel) for simulating doping or just displacing the chemical potential for the calculations in the undoped compound (upper panel). The values of doping chosen are, as explained above, in the range where the AF ordering survives and our calculations are more reliable, and also the interval where it was predicted the enhancement in TE figure of merit with doping for the unstrained compound.\cite{la2nio4_therm_prb} It was seen in previous studies\cite{la2nio4_Seebeck_delta} that large thermopower occurs for $\delta$= 0 and then gets rapidly reduced as doping is introduced. Here we can analyze in more detail the effects of such doping levels in the thermopower together with the effect of strain.

Looking at Fig. \ref{S_delta}, one can see that in the tensile strain region (a $>$ 3.90 \AA) the thermopower is enhanced, whereas in the compressive strain region it is reduced. This is the expected result because, as we have discussed above, enlarging the xy-plane leads to a wider gap and hence the thermopower will increase. We see that, as doping is increased, things get slightly more complicated than that, at the same time that thermopower is reduced. Putting everything together, values of the thermopower at room temperature in the range of S $\sim$ 150 $\mu$V/K are predicted up to $\delta$= 0.10 in the tensile strain regime, and up to 200 $\mu$V/K for $\delta$= 0.05. We observe in the lower panel of Fig. \ref{S_delta} that the results using the VCA differ substantially in quantitative terms (some vague trends are maintained, but the values are very different to those obtained shifting the chemical potential in the undoped compound). We argue that using the VCA here, for these substantial doping levels, is necessary to yield a correct description of the system. We can see that the two methods give more similar accounts at low-doping ($\delta$ $\sim$ 0.05), where shifting the chemical potential in the calculation for the undoped compound is enough to describe more or less accurately the system (it will work even better for $\delta$ $<$ 0.05), but going to larger doping levels, one needs to include the change in the overall potential caused by the oxygen excess in the system to get a good description. In the VCA-based calculations, we see that as doping increases, the values of the thermopower decay quite fast, except at high temperatures. But even at high temperatures, the lattice parameter window where the response is promising becomes narrower, around 4.00 \AA. Focusing on that tensile strain value, we see that the values of the thermopower at room temperature are not promising: 75 $\mu$V/K at $\delta$= 0.05, 40 $\mu$V/K at $\delta$= 0.10 and only 25 $\mu$V/K at $\delta$= 0.15, yet the high temperature response becomes enhanced. It has been observed experimentally that La$_2$NiO$_{4+\delta}$ has an increasing thermopower at high temperatures.\cite{la2nio4_S_1,la2nio4_S_3}

\begin{figure*}[ht]
\begin{center}
\includegraphics[width=0.32\textwidth,draft=false]{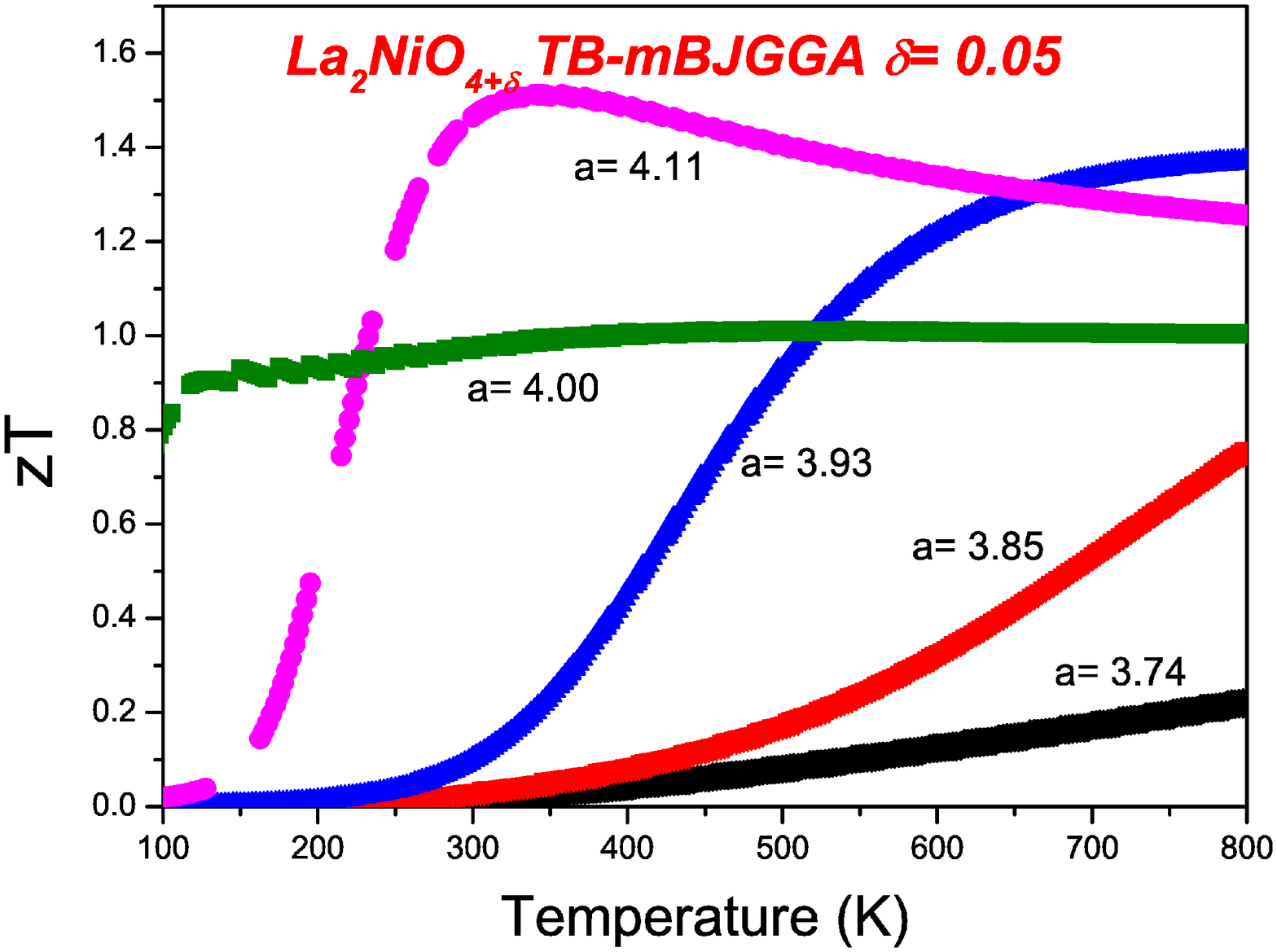}
\includegraphics[width=0.32\textwidth,draft=false]{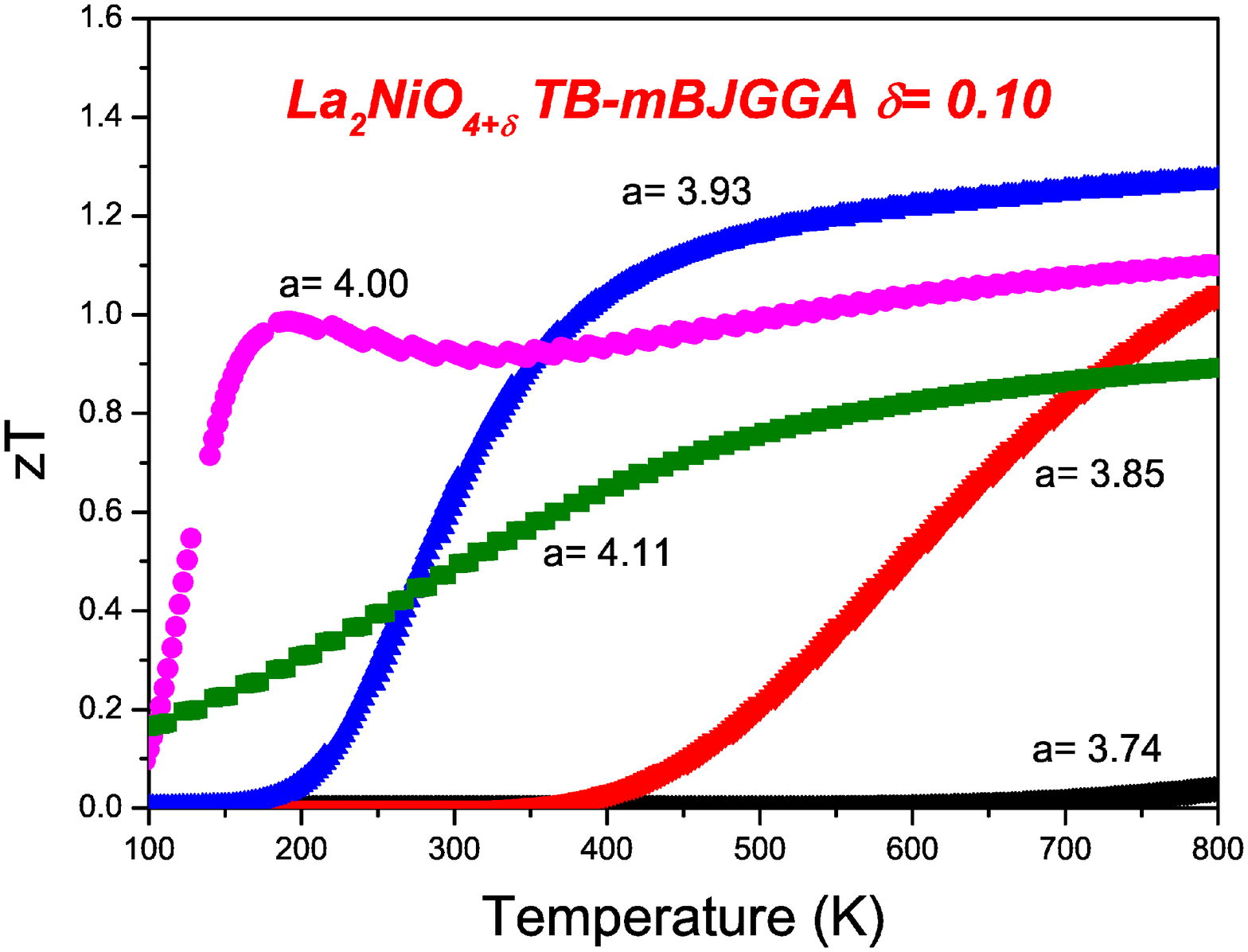}
\includegraphics[width=0.32\textwidth,draft=false]{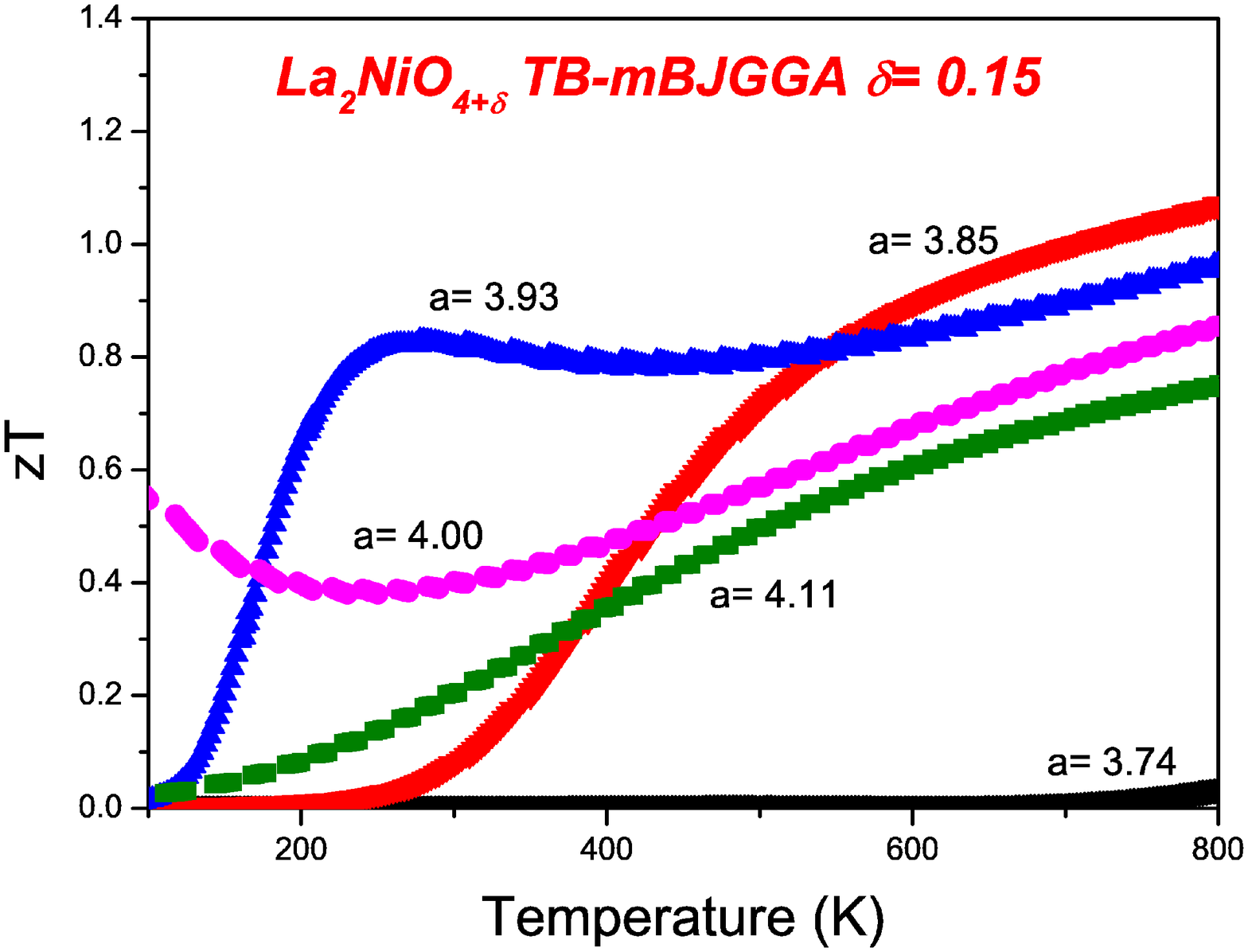}
\includegraphics[width=0.32\textwidth,draft=false]{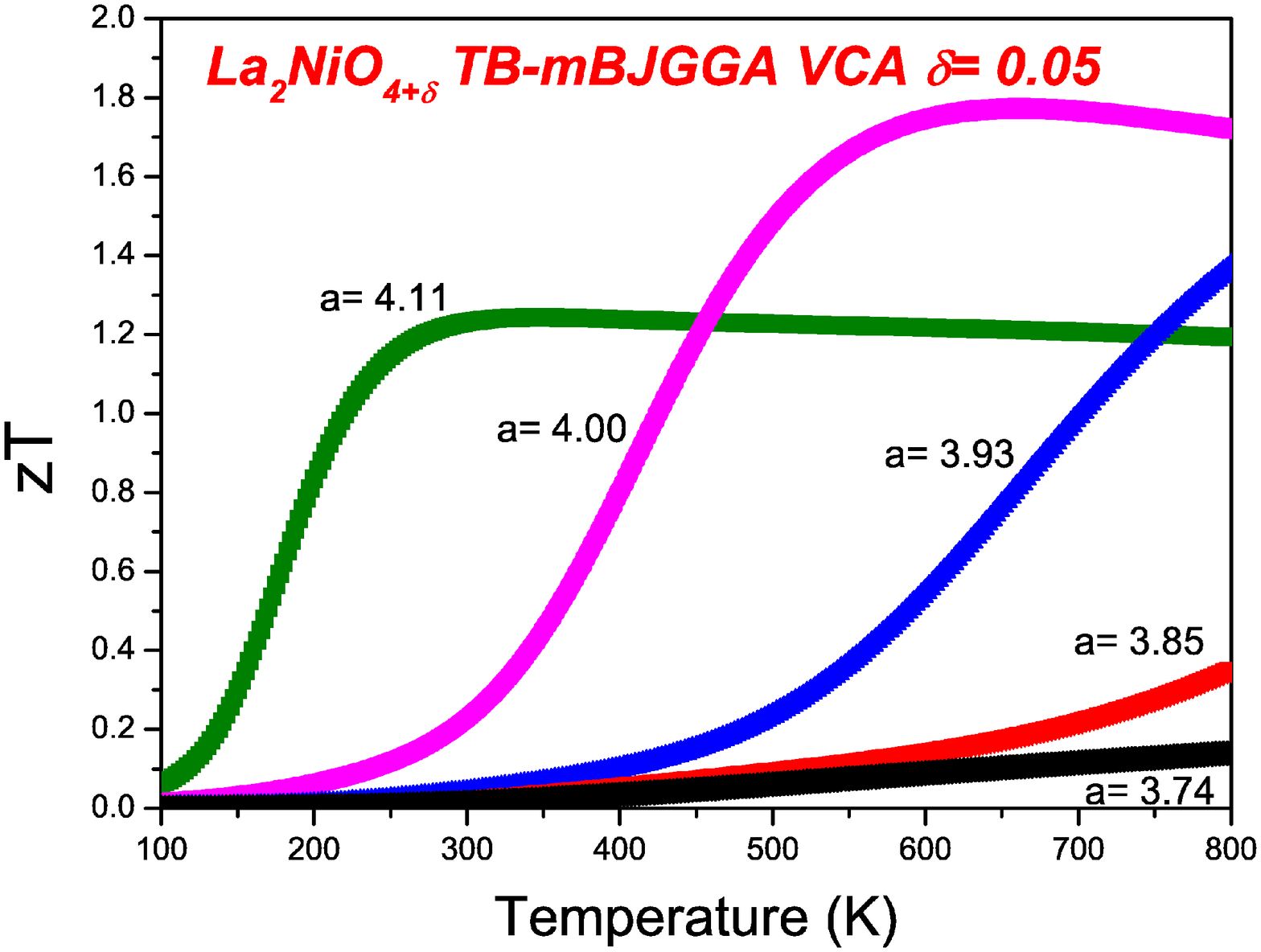}
\includegraphics[width=0.32\textwidth,draft=false]{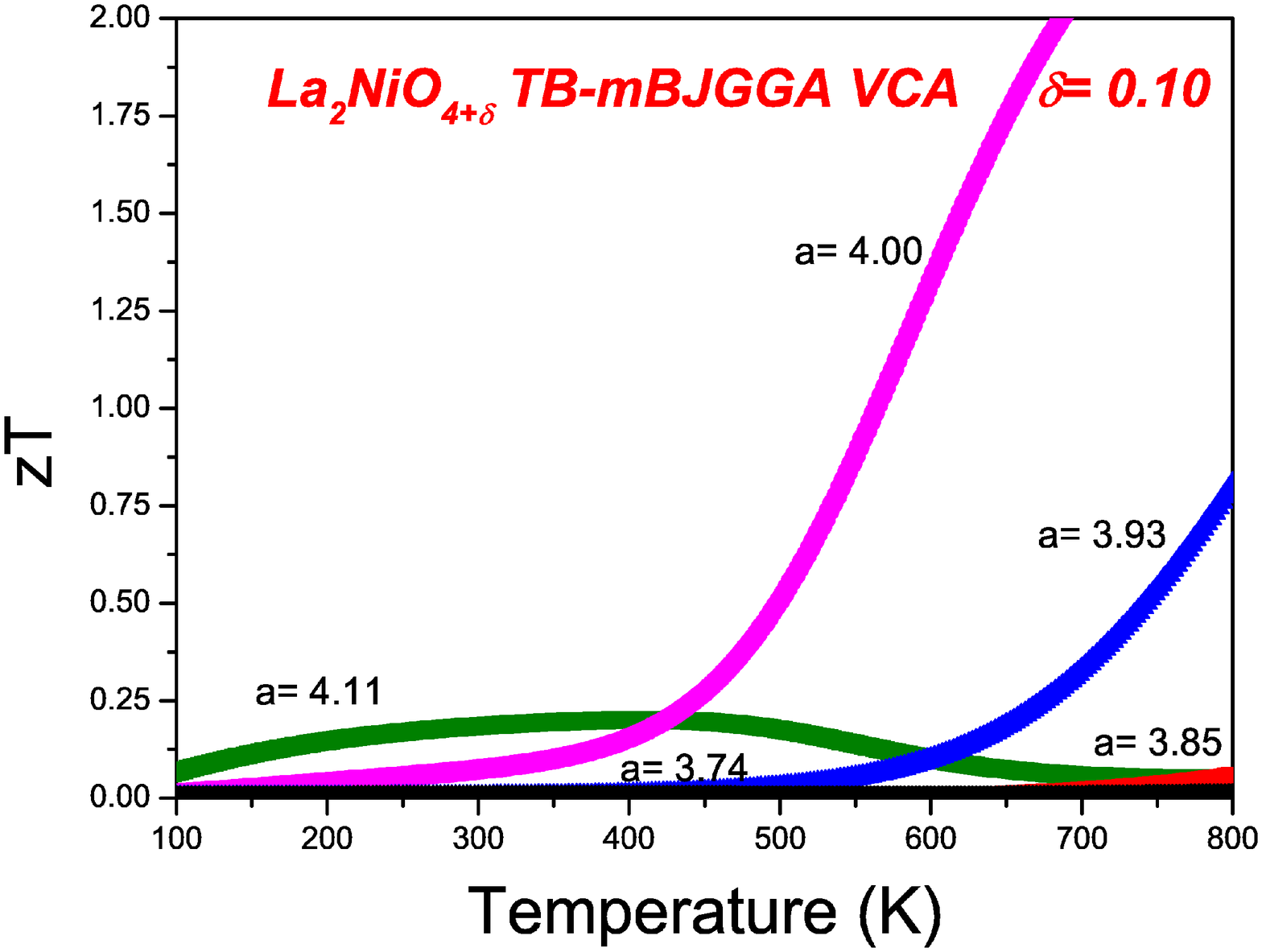}
\includegraphics[width=0.32\textwidth,draft=false]{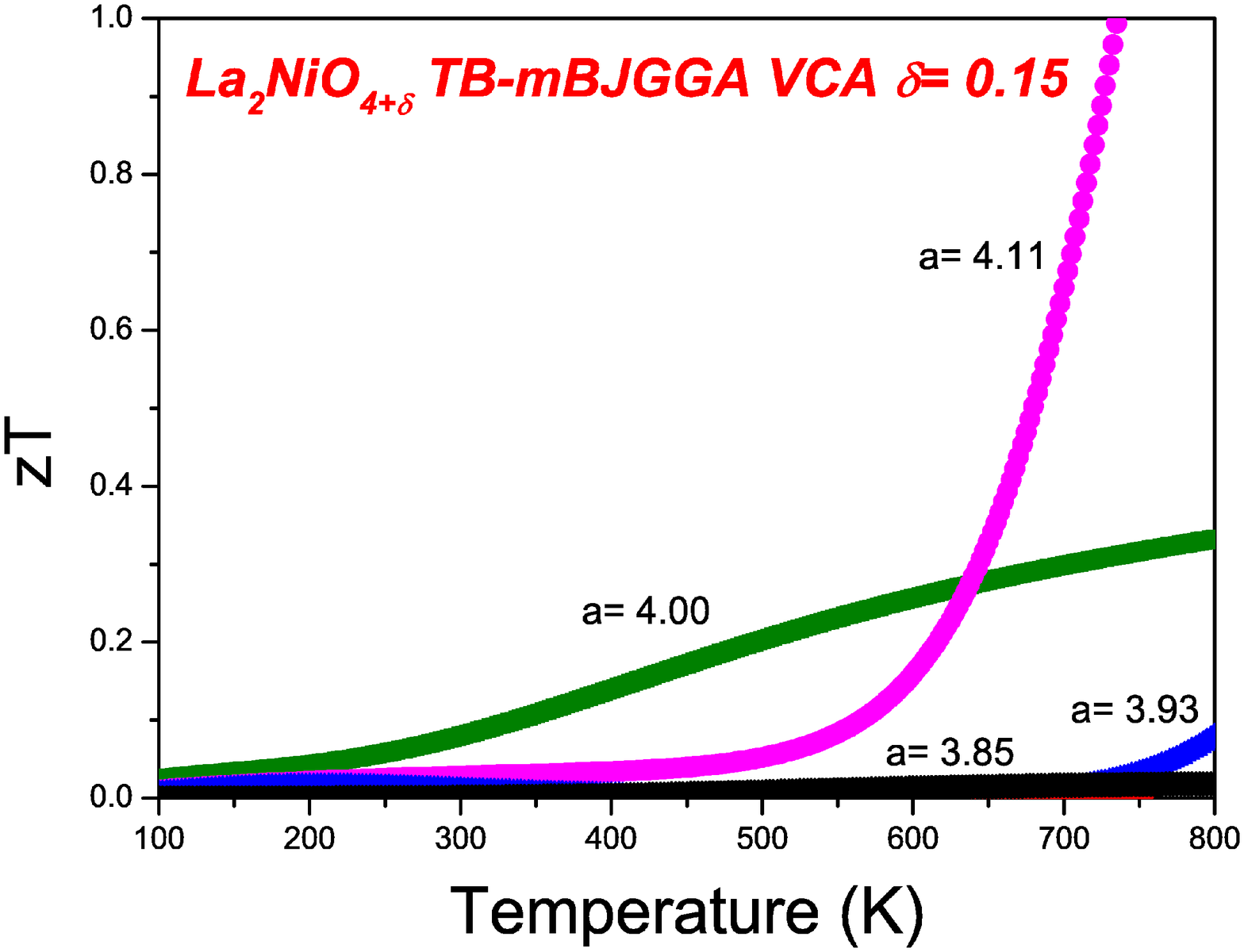}
\caption{(Color online.)  Figure of merit of La$_2$NiO$_{4+\delta}$ as a function of temperature for three doping values: $\delta$= 0.05, $\delta$= 0.10 and $\delta$= 0.15 for various values of the in-plane lattice parameter (different strains). The upper was calculated for an undoped La$_2$NiO$_4$ shifting the chemical potential to the desired carrier concentration, whereas the lower panel was calculated using the VCA. The latter yields a more reliable and accurate description of the system. The exchange-correlation potential used for calculating all the transport properties was TB-mBJGGA. Observe that for various lattice parameters in the doping range $\delta$= 0.05 - 0.10, the electronic figure of merit exceeds unity. }\label{zt_delta}
\end{center}
\end{figure*}

However, we know that the larger the $a$ parameter is, the wider the band gap gets (and that is why even upon doping large thermopower values can be obtained) but the conductivity will be reduced (in-plane hopping decreases). We can compile both effects in the calculation of the dimensionless TE figure of merit, that is presented in Fig. \ref{zt_delta} for several doping levels and lattice parameters (the same ones we have used for the thermopower) and with the same calculation schemes described above. We see at low-doping, around $\delta$= 0.05 and for a slight tensile strain ($a$ $>$ 3.90 \AA), values of the electronic-only TE figure of merit exceeding unity can be obtained even at room temperature for the larger lattice parameters tested, and at high temperatures for a very small tensile strain. When moving the system towards the compressive strain limit, even though in principle it should increase the conductivity, the gain through that term does not win over the loss in thermopower (which is substantial, as we saw above). We can see again that neglecting VCA and just shifting the chemical potential in the undoped compound will produce a large overestimation of the TE figure of merit (but the agreement with VCA is again better at small doping). As doping is increased, as we saw for the thermopower, a promising TE response only occurs at large temperature and for a very critical range of in-plane lattice parameters (around $a$= 4.00 \AA\ for $\delta$= 0.10 and $a=$ 4.11 \AA\ for $\delta$= 0.15). This means that one has to optimize doping for each strain analyzed to obtain a significant TE figure of merit. For the larger tensile strain considered (4.11 \AA), the behavior is quite peculiar, not being monotonic with doping. Similar non-monotonic behavior is observed also for the thermopower. As $a$ increases, the $x^2-y^2$ band gets lower in energy relative to the $xz/yz$ bands, and eventually it gets so close to them that hole-doping introduces states of those bands very close to the Fermi level. The relevant contributions from those many bands make the behavior and temperature evolution in that case much more complicated to analyze and understand. The large lattice parameter window with electronic-only figure of merit exceeding unity in the tensile strain limit only occurs for small values of doping around $\delta$= 0.05.

As we have been arguing, one possible route to reduce the thermal conductivity is the growth in thin films. It has been recently observed how oxygen non-stoichiometries can lead to a significant reduction of the thermal conductivity by creating new scattering centers that would hamper phonon propagation in SrTiO$_{3-\delta}$ thin films at very low doping levels.\cite{fran_sto,ramesh_sto} Experiments are required to understand if the same mechanism is at play in La$_2$NiO$_{4+\delta}$. If that is the case, the increase in the electronic figure of merit via hole-doping could be a promising route to an enhanced TE performance in La$_2$NiO$_{4+\delta}$ thin films. Moreover, the growth in the form of thin films will increase the electronic conductivity as has been shown in the past,\cite{la2nio4_films_sigma} due to the larger weight of the in-plane conduction (the largely layered structure impedes conductivity in the off-plane direction). Also, a thin film geometry is expected to reduce thermal conductivity by preventing the propagation of long-wavelength phonons, apart from the possible reduction imposed by vacancy-related scattering. As we have discussed before, effects related to vacancy localization beyond the average treatment of the dopants we are considering here (both using VCA and a shift in the chemical potential) are beyond the scope of this work.
The recipe, according to our calculations, would be the growth of La$_2$NiO$_{4+\delta}$ thin films with oxygen excess on the order of $\delta$ $\sim$ 0.10, grown on top of a substrate that provides some tensile strain, and studied at high temperatures, where the most plausible applications of this oxide and the larger figure of merit will occur.

To further validate our results, we can give an estimate of the overall $zT$ (up to now we have studied the electronic part only). Considering the Wiedemann-Franz relation, the figure of merit can be rewritten as $zT$= $\sigma$TS$^2$/$\kappa$= $\kappa_{e}$$S^2$/$\kappa$$L_0$, $L_0$ being the Lorenz number with a value for free electrons $L_0$= 2.45 x 10$^{-8}$ $(V/K)^2$ and $\kappa$ $<$ $\kappa_{e}$ in general, but approaching 1 at large doping. At high enough temperatures, the lattice term typically decreases as $1/T$. This type of behavior can be observed in Ref. \onlinecite{la2nio4_kappa}. Taking the thermal conductivity experimental values for the unstrained compound at the transition temperature from Ref.  \onlinecite{la2nio4_kappa}, the electrical conductivity from Ref. \onlinecite{la2nio4_films_sigma}, and using our calculated values for the thermpower, we can obtain an estimation of the values of the overall $zT$. At 400 K, for $a$= 4.11 \AA, a $zT$ of 0.3 can be achieved for $\delta$= 0.05 while for $\delta$= 0.15 only a figure of merit of 0.04 can be obtained. For $\delta$= 0.10 and $a$= 4.00 \AA, a lower value of 0.02 is estimated at 400 K. However, at high temperatures (above 700 K), $zT$s close to one ($\sim$ 0.8) can be obtained for $a$= 4.00 \AA\ and values of doping between 0.05 and 0.10. These $zT$ values are already very promising but if, in addition, the thermal conductivity of the system can be reduced with respect to the bulk value when grown in thin-film geometry, our calculations suggest that values of $zT$ beyond one could in principle be achieved.



\section{Summary}\label{summary}

Our ab initio calculations for the compound indicate some promising features in hole-doped La$_2$NiO$_{4+\delta}$ as a possible oxide thermoelectric material. If the thermal conductivity can be reduced by nanostructuring, e.g. in the form of thin films, the system could show enhanced thermoelectric performance at low hole-doping levels, attainable by the appropriate control of the apical oxygen content. Our calculations show that a region with relatively large Seebeck coefficient exists in this compound at small doping levels and the appropriate tensile strain, within the realistic AF description. A careful experimental study needs to be performed in this respect controlling the oxygen content, and also making thin films with the appropriate oxygen composition. Hole-doping will increase the electrical conductivity, as the thin film geometry does, which together with the reduction in thermal conductivity could leave room for an improvement of the thermoelectric figure of merit of this layered nickelate. We suggest to explore the region of tensile strain, growing on substrates with a lattice parameter slightly larger than that of SrTiO$_3$ (about 3.95 - 4.05 \AA). Our calculations predict a reasonable thermoelectic performance (with figure of merit close to 1, $zT$ $\sim$ 0.8) in those conditions of doping and tensile strain at high temperatures.

\section{Acknowledgments}

We thank F. Rivadulla and C.X. Quintela for fruitful discussions. V.P. and A.S.B. acknowledge the Spanish Government for support through the Ram\'on y Cajal program and FPU program, respectively. Financial support was given from the Ministry of Science of Spain through project MAT-200908165.


\end{document}